\documentclass[nofootinbib,floatfix,twocolumn,superscriptaddress,a4paper,prl]{revtex4}

\usepackage{amsmath}
\usepackage{amssymb}
\usepackage{graphicx}
\usepackage[latin1]{inputenc}
\usepackage{mathrsfs}
\usepackage{xspace}
\usepackage{verbatim}
\usepackage[dvips]{color}

\newcommand{\Br}{\text{Br}}

\newcommand{\SARAH}{{\tt SARAH}\xspace}
\newcommand{\SPheno}{{\tt SPheno}\xspace}
\newcommand{\HB}{{\tt HiggsBounds}\xspace}
\newcommand{\HS}{{\tt HiggsSignals}\xspace}
\newcommand{\SSP}{{\tt SSP}\xspace}
\newcommand{\flavourkit}{{\tt FlavourKit}\xspace}

\newcommand{\BL}{{\ensuremath{B-L}}\xspace}
\newcommand{\UBL}{{\ensuremath{U(1)_{B-L}}}\xspace}
\newcommand{\BLSSM}{BLSSM\xspace}

\newcommand{\vev}{\textit{vev}\xspace}
\newcommand{\vevs}{\textit{vev}s\xspace}

\newcommand{\EQ}{eq.\xspace}
\newcommand{\EQS}{eqs.\xspace}

\newcommand{\FIG}{Fig.\xspace}

\newcommand{\EG}{\textit{e.g.}\xspace}
\newcommand{\IE}{\textit{i.e}.\xspace}
\newcommand{\REF}{Ref.\xspace}

\def\gsim{\raise0.3ex\hbox{$\;>$\kern-0.75em\raise-1.1ex\hbox{$\sim\;$}}}

\newcommand{\gBL}[1]{{\ensuremath{g_{BL}^{#1}}}\xspace}
\newcommand{\gmix}{{\ensuremath{\bar{g}}}\xspace}
\newcommand{\gsqsum}{{\ensuremath{g_{\Sigma}^{2}}}\xspace}
\newcommand{\rsnu}{R-sneutrino\xspace}
\newcommand{\rsnus}{R-sneutrinos\xspace}

\newcommand{\lsnus}{L-sneutrinos\xspace}

\newcommand{\mZp}{{\ensuremath{M_{Z^{\prime}}}}\xspace}
\newcommand{\mZpSq}{{\ensuremath{M_{Z^{\prime}}^{2}}}\xspace}
\newcommand{\mReSnuSq}[0]{{\ensuremath{m^{2}_{{\tilde{\nu}}^{S}}}}\xspace}
\newcommand{\mImSnuSq}[0]{{\ensuremath{m^{2}_{{\tilde{\nu}}^{P}}}}\xspace}

\begin{document}

\date{\today}

\title{The Higgs sector of the minimal SUSY $B-L$ model}

\author{Lorenzo~Basso}\email[E-mail: ]{Lorenzo.Basso@iphc.cnrs.fr}
\affiliation{Universit\`e de Strasbourg, IPHC, 23 rue du Loess 67037 Strasbourg, France}
\affiliation{CNRS, UMR7178, 67037 Strasbourg, France}

\begin{abstract}
\noindent
I review the Higgs sector of the $U(1)_{B-L}$ extension of the minimal supersymmetric standard model (MSSM). I will show that the gauge kinetic mixing plays a crucial role in the Higgs phenomenology. Two light bosons are present, a MSSM-like one and a $B-L$-like one, that mix at one loop solely due to the gauge mixing. After briefly looking at constraints from flavour observables, new decay channels involving right-handed (s)neutrinos are presented. Finally, it will be reviewed how model features pertaining to the gauge extension affect the model phenomenology, concerning the existence of R-Parity-conserving minima at loop level and the Higgs-to-diphoton coupling.
\end{abstract}

\maketitle

\section{Introduction}
The recently discovered Higgs boson is considered as the last missing piece of the standard model (SM) of particle physics. Nonetheless, several firm observations univocally call for its extension. Mainly but not limited to, the presence of dark matter, the neutrino masses and mixing pattern, the stability of the SM vacuum, the hierarchy problem. Supersymmetry (SUSY) has long been considered as the most appealing framework to extend the SM. Its minimal realisations (MSSM and its constrained versions~\footnote{For a review, see Ref.\cite{Ellis:2012nv}.}) start however to feel considerable pressure to accommodate the recent findings, especially the measured Higgs mass of $125$ GeV. Despite not in open contrast with the MSSM, the degree of fine tuning required to achieve it is more and more felt as unnatural. In order to alleviate this tension, non-minimal SUSY realisations can be considered. One can either extend the MSSM by the inclusion of extra singlets (\EG NMSSM~\cite{Ellwanger:2009dp}) or by extending its gauge group.
Concerning the latter, one of the simplest
possibilities  is to add
an additional Abelian gauge group. I will focus here on the presence
of an $U(1)_{B-L}$ group which can be a result of an \(E_8 \times
E_8\) heterotic string theory (and hence M-theory)
\cite{Buchmuller:2006ik,Ambroso:2009sc,Ambroso:2010pe}. 
This model, the minimal $R$-parity-conserving \BL supersymmetric
 standard model (\BLSSM in short),
 was proposed in
\cite{Khalil:2007dr,FileviezPerez:2010ek} and neutrino masses are
obtained via a type I seesaw mechanism. Furthermore, it could
 help  to understand the origin of $R$-parity and its
 possible spontaneous violation in supersymmetric models
\cite{Khalil:2007dr,Barger:2008wn,FileviezPerez:2010ek}
 as well as the mechanism of leptogenesis \cite{Pelto:2010vq,Babu:2009pi}.

It was early pointed out that the presence of two Abelian
gauge groups in this model gives rise to kinetic mixing terms of the
form
\begin{equation}
\label{eq:offfieldstrength}
- \chi_{ab}  \hat{F}^{a, \mu \nu} \hat{F}^b_{\mu \nu}, \quad a \neq b
\end{equation}
that are allowed by gauge and Lorentz invariance \cite{Holdom:1985ag},
as $\hat{F}^{a, \mu \nu}$ and $\hat{F}^{b, \mu \nu}$ are
 gauge-invariant quantities by themselves,
 see \EG \cite{Babu:1997st}. Even
if these terms are absent at tree level at a particular scale, they
will in general be generated by RGE effects
\cite{delAguila:1988jz,delAguila:1987st}.  These terms can have a
sizable effect on the mass spectrum of this model, as studied in detail in \REF~\cite{O'Leary:2011yq}, and on the dark matter, where several
scenarios would not work if it is neglected, as thoroughly investigated in \REF~\cite{Basso:2012gz}.
In this work, I will review the properties of the Higgs sector of the model. Two light states exist, a MSSM-like boson and a $B-L$-like boson. After reviewing the model, I will show that a large portion of parameter space exists where the SM-like Higgs boson has a mass compatible with its measure, both in a ``normal'' ($M_{H_2} > M_{H_1}=125$ GeV) and in a ``inverted'' hierarchy ($M_{H_1} < M_{H_2}=125$ GeV), also in agreement with bounds from low energy observables and dark matter relic abundance. The phenomenological properties of the two lightest Higgs bosons will be systematically investigated, where once again the gauge mixing is shown to be fundamental.
The presence of extra D-terms arising from the new $U(1)_{B-L}$ sector, as compared to models based on the SM gauge symmetry, has a large impact on the model phenomenology. They affect both the vacuum structure of the model and the Higgs sector, in particular enhancing the Higgs-to-diphoton coupling. Both these issues will be reviewed here, despite the latter is disfavoured by recent data~\cite{Khachatryan:2014jba}, to show model features beyond the MSSM.

\section{The model}
\label{sec:model}
For a detailed discussion of the masses of
all particles as well as of the corresponding one-loop corrections we
refer to \cite{O'Leary:2011yq}. Attention will be payed on the main aspects
of the $U(1)$ kinetic mixing since it has important consequence for
the scalar sector.
For the numerical investigations that will be shown, we used the \SPheno version \cite{Porod:2003um,Porod:2011nf} created with \SARAH \cite{Staub:2008uz,Staub:2009bi,Staub:2010jh,Staub:2012pb,Staub:2013tta} for the \BLSSM. For the standardised model definitions, see \REF~\cite{Basso:2012ew}, while for a review of the model implementation in \SARAH, see \REF~\cite{Staub:2015kfa}. This spectrum calculator performs a two-loop RGE evaluation and calculates the mass spectrum at one loop. In addition, it calculates the decay widths and branching ratios (BRs) of all SUSY and Higgs particles as well as low-energy observables like $(g-2)_\mu$. We will discuss the most constrained scenario with a universal scalar mass $m_0$, a universal gaugino mass $M_{1/2}$ and trilinear soft-breaking couplings proportional to the superpotential coupling ($T_i = A_0 Y_i$) at the GUT scale. Other input parameters are $\tan\beta$, $\tan\beta'$, $M_{Z'}$, $Y_x$, and $Y_\nu$. They will be defined in the following section. The numerical study here presented has been performed by randomly scanning over the independent input parameters above described via the \SSP toolbox~\cite{Staub:2011dp}, while low energy observables such as BR($\mu\to e\gamma$) and BR($\mu\to 3e$) have been evaluated with the \flavourkit package~\cite{Porod:2014xia}. Furthermore, during the scans all points have been checked with \HB-4.1.1~\cite{Bechtle:2008jh,Bechtle:2011sb,Bechtle:2013gu,Bechtle:2013wla}, both in the ``normal'' hierarchy and in the ``inverted`` hierarchy case.

\subsection{Particle content and superpotential}
The model consists of three generations of matter particles including
right-handed neutrinos which can, for example, be embedded in $SO(10)$
16-plets. Moreover, below the GUT scale the usual MSSM Higgs doublets
are present as well as two fields $\eta$ and $\bar{\eta}$ responsible
for the breaking of the \UBL. The $\eta$ field is also responsible for
generating a Majorana mass term for the right-handed neutrinos and
thus we interpret its \BL charge as its lepton number.
Likewise is for $\bar{\eta}$, and we call these fields bileptons since
they carry twice the lepton number of (anti-)neutrinos.  The quantum numbers of the chiral superfields with respect to $U(1)_Y
\times SU(2)_L \times SU(3)_C \times \UBL$ are summarised in Table~\ref{tab:cSF}.
\begin{table} 
\centering
\begin{tabular}{|c|c|c|c|c|c|} 
\hline \hline 
Superfield & Spin 0 & Spin \(\frac{1}{2}\) & Generations & $G_{SM}\otimes\, \UBL$ \\ 
\hline 
\(\hat{Q}\) & \(\tilde{Q}\) & \(Q\) & 3
 & \((\frac{1}{6},{\bf 2},{\bf 3},\frac{1}{6}) \) \\ 
\(\hat{d}^c\) & \(\tilde{d}^c\) & \(d^c\) & 3
 & \((\frac{1}{3},{\bf 1},{\bf \overline{3}},-\frac{1}{6}) \) \\ 
\(\hat{u}^c\) & \(\tilde{u}^c\) & \(u^c\) & 3
 & \((-\frac{2}{3},{\bf 1},{\bf \overline{3}},-\frac{1}{6}) \) \\ 
\(\hat{L}\) & \(\tilde{L}\) & \(L\) & 3
 & \((-\frac{1}{2},{\bf 2},{\bf 1},-\frac{1}{2}) \) \\ 
\(\hat{e}^c\) & \(\tilde{e}^c\) & \(e^c\) & 3
 & \((1,{\bf 1},{\bf 1},\frac{1}{2}) \) \\ 
\(\hat{\nu}^c\) & \(\tilde{\nu}^c\) & \(\nu^c\) & 3
 & \((0,{\bf 1},{\bf 1},\frac{1}{2}) \) \\ 
\(\hat{H}_d\) & \(H_d\) & \(\tilde{H}_d\) & 1
 & \((-\frac{1}{2},{\bf 2},{\bf 1},0) \) \\ 
\(\hat{H}_u\) & \(H_u\) & \(\tilde{H}_u\) & 1
 & \((\frac{1}{2},{\bf 2},{\bf 1},0) \) \\ 
\(\hat{\eta}\) & \(\eta\) & \(\tilde{\eta}\) & 1
 & \((0,{\bf 1},{\bf 1},-1) \) \\ 
\(\hat{\bar{\eta}}\) & \(\bar{\eta}\) & \(\tilde{\bar{\eta}}\) & 1
 & \((0,{\bf 1},{\bf 1},1) \) \\ 
\hline \hline
\end{tabular} 
\caption{Chiral superfields and their quantum numbers under $G_{SM}\otimes\, \UBL$, where $G_{SM} = $ \((U(1)_Y\otimes\, SU(2)_L\otimes\, SU(3)_C)\) .}
\label{tab:cSF}
\end{table}

The superpotential is given by
\begin{eqnarray} 
\nonumber 
W & = & \, Y^{ij}_u\,\hat{u}^c_i\,\hat{Q}_j\,\hat{H}_u\,
- Y_d^{ij} \,\hat{d}^c_i\,\hat{Q}_j\,\hat{H}_d\,
- Y^{ij}_e \,\hat{e}^c_i\,\hat{L}_j\,\hat{H}_d\, \\ \nonumber
 & & +\mu\,\hat{H}_u\,\hat{H}_d\,
+Y^{ij}_{\nu}\,\hat{\nu}^c_i\,\hat{L}_j\,\hat{H}_u\,
- \mu' \, \hat{\eta}\,\hat{\bar{\eta}}\,
+Y^{ij}_x\,\hat{\nu}^c_i\,\hat{\eta}\,\hat{\nu}^c_j\, \\
\label{eq:superpot}
 & &
\end{eqnarray} 
and we have the additional soft SUSY-breaking terms:
\begin{eqnarray}
\nonumber \mathscr{L}_{SB} &= & \mathscr{L}_{MSSM}
 - \lambda_{\tilde{B}} \lambda_{\tilde{B}'} {M}_{B B'} 
 - \frac{1}{2} \lambda_{\tilde{B}'} \lambda_{\tilde{B}'} {M}_{B'}\\ \nonumber
&&  - m_{\eta}^2 |\eta|^2 - m_{\bar{\eta}}^2 |\bar{\eta}|^2 
 - {m_{\nu^c,ij}^{2}} (\tilde{\nu}_i^c)^* \tilde{\nu}_j^c \\
&& - \eta \bar{\eta} B_{\mu'} + T^{ij}_{\nu}  H_u \tilde{\nu}_i^c \tilde{L}_j
 + T^{ij}_{x} \eta \tilde{\nu}_i^c \tilde{\nu}_j^c 
\end{eqnarray}
$i,j$ are generation indices. Without loss of generality one can take
$B_\mu$ and $B_{\mu'}$ to be real. The extended gauge group breaks to
$SU(3)_C \otimes U(1)_{em}$ as the Higgs fields and bileptons receive
vacuum expectation values (\vevs):
\begin{eqnarray*} 
H_d^0 = & \, \frac{1}{\sqrt{2}} \left(\sigma_{d} + v_d  + i \phi_{d} \right),
\hspace{1cm}
H_u^0 = \, \frac{1}{\sqrt{2}} \left(\sigma_{u} + v_u  + i \phi_{u} \right)\\ 
\eta
= & \, \frac{1}{\sqrt{2}} \left(\sigma_\eta + v_{\eta} + i \phi_{\eta} \right),
\hspace{1cm}
\bar{\eta}
= \, \frac{1}{\sqrt{2}} \left(\sigma_{\bar{\eta}} + v_{\bar{\eta}}
 + i \phi_{\bar{\eta}} \right)
\end{eqnarray*} 
We define $\tan\beta' = v_{\eta}/v_{\bar{\eta}}$
 in analogy to
the ratio of the MSSM \vevs ($\tan\beta = v_{u}/v_{d}$).

\subsection{Gauge kinetic mixing}
\label{subsec:kineticmixing}
As already mentioned in the introduction, the presence of two Abelian
gauge groups in combination with the given particle content gives 
rise to a new
effect absent in any model with just one Abelian
gauge group:  gauge kinetic mixing. This can be seen most
 easily by
inspecting the matrix of the
anomalous dimension, which for our model at one loop reads
\begin{equation}
\label{eq:gammaMatrix}
 \gamma = \frac{1}{16 \pi^2}
  \left( \begin{array}{cc} \frac{33}{5} & 6 \sqrt{\frac{2}{5}} \\
   6 \sqrt{\frac{2}{5}} & 9 \end{array} \right) ,
\end{equation}
with typical GUT normalisation of the two Abelian gauge groups, \IE  \(\sqrt{{3/5}}\) for
\(U(1)_{Y}\) and \(\sqrt{{3/2}}\) for \UBL~\cite{FileviezPerez:2010ek}.
Therefore, even if at the GUT scale the $U(1)$ kinetic mixing terms
are zero, they are induced via RGE evaluation at lower scales. 
It turns out that it is more convenient to work with non-canonical
covariant derivatives rather than with off-diagonal field-strength tensors
 as in \EQ~(\ref{eq:offfieldstrength}).  However, both approaches
are equivalent \cite{Fonseca:2011vn}. Therefore, in the following, we
consider covariant derivatives of the form
$\displaystyle D_\mu  = \partial_\mu - i Q_{\phi}^{T} G  A $
where \(Q_{\phi}\) is a vector containing the charges of the field $\phi$ with
respect to the two Abelian gauge groups, $G$ is the gauge coupling matrix
\begin{equation}
 G = \left( \begin{array}{cc} g_{YY} & g_{YB} \\
                              g_{BY} & g_{BB} \end{array} \right)
\end{equation}
and $A$ contains the gauge bosons $A = ( A^Y_\mu, A^B_\mu )^T$.

As long as the two Abelian gauge groups are unbroken, we have still
the freedom to perform a change of basis by means of a suitable rotation.
A convenient choice is the basis where \(g_{B Y}=0\), since in this case
only the Higgs doublets contribute to the gauge boson
mass matrix of the $SU(2)_L \otimes U(1)_Y$ sector, while the impact of
$\eta$ and $\bar{\eta}$ is only in the off-diagonal elements.
Therefore we choose the following basis at the electroweak scale
\cite{Chankowski:2006jk}:
\begin{eqnarray}
\label{eq:gYYp}
 g'_{Y Y}
 = & \frac{g_{YY} g_{B B} - g_{Y B} g_{B Y}}{\sqrt{g_{B B}^2 + g_{B Y}^2}}
 = g_1  \\
 g'_{B B} = & \sqrt{g_{B B}^2 + g_{B Y}^2} = \gBL{} \\
 \label{eq:gtilde}
 g'_{Y B}
 = & \frac{g_{Y B} g_{B B} + g_{B Y} g_{YY}}{\sqrt{g_{B B}^2 + g_{B Y}^2}}
 = \gmix \\
 g'_{B Y} = & 0
\label{eq:gBYp}
\end{eqnarray}

When unification at some large scale ($\sim 2 \cdot 10^{16}$ GeV) is imposed, \IE, $g_1^{GUT}=g_2^{GUT}=\gBL{}^{GUT}$ and $g^{\prime\, (GUT)}_{Y B}=g^{\prime\, (GUT)}_{B Y} =  0$, at SUSY scale we get~\cite{O'Leary:2011yq}
\begin{eqnarray} \label{eq:gBLsusy}
\gBL{} &=& 0.548\, ,\\ \label{eq:gtildesusy}
\bar{g} &\simeq& -0.147\, .
\end{eqnarray}

\subsection{Tadpole equations}
\label{subsec:tadpoles}
The minimisation of the scalar potential is here described in the so-called tadpole method. We can solve the tree-level tadpole equations arising from the minimum conditions
of the vacuum with respect to \(\mu, B_\mu, \mu'\) and \(B_{\mu'}\).
Using $v_x^2=v_{\eta}^{2} + v_{\bar{\eta}}^{2}$ and 
$v^2=v_{d}^{2}+ v_{u}^{2}$ we obtain 
\begin{widetext}
\begin{eqnarray}
\label{eq:tadmu}
 |\mu|^2 = & \frac{1}{8} \Big(\Big(2 \gmix \gBL{} v_x^{2}
 \cos(2 {\beta'})
    -4 m_{H_d}^2  + 4 m_{H_u}^2 \Big)\sec(2 \beta)
    -4 \Big(m_{H_d}^2 + m_{H_u}^2\Big)
 - \Big(g_{1}^{2} + {\gmix}^{2} + g_{2}^{2}\Big)v^{2} \Big)\\ \label{eq:tadBmu}
 B_\mu =&-\frac{1}{8} \Big(-2 \gmix \gBL{} v_x^{2}
 \cos(2 {\beta'})
    + 4 m_{H_d}^2  -4 m_{H_u}^2
  + \Big(g_{1}^{2} + {\gmix}^{2} + g_{2}^{2}\Big)v^{2} \cos(2 \beta)
   \Big)\tan(2 \beta )     \\
 |\mu'|^2 =& \frac{1}{4} \Big(-2 \Big( \gBL{2} v_x^{2}
  + m_{\eta}^2 + m_{\bar{\eta}}^2\Big) + \Big(2 m_{\eta}^2 - 2 m_{\bar{\eta}}^2
  + \gmix \gBL{} v^{2} \cos(2 \beta) \Big)
\sec(2 {\beta'}) \Big) \\
\label{eq:tadBmuP}
 B_{\mu'} =&  \frac{1}{4} \Big(-2 \gBL{2} v_x^{2} \cos(2 {\beta'}) 
   + 2 m_{\eta}^2  -2 m_{\bar{\eta}}^2
  + \gmix \gBL{} v^{2} \cos(2 \beta)
   \Big) \tan(2 {\beta'} )
\end{eqnarray}
\end{widetext}
$\mZp \simeq \gBL{} v_x$  and, thus, we find an approximate
relation between $\mZp$ and $\mu'$
\begin{eqnarray} \nonumber
 \mZp^2 &\simeq &
 - 2 |\mu'|^2\\ \nonumber
 && + \frac{4 (m_{\bar{\eta}}^2 - m_{\eta}^2 \tan^2 \beta')
- v^2 \gmix \gBL{} \cos\beta (1+\tan\beta') }{2 (\tan^2 \beta'
 - 1) }\\
 && \label{eq:tadpole_MZp}
\end{eqnarray}
For the numerical results, the one-loop corrected equations are used, which lead to a shift of the solutions in eqs.~(\ref{eq:tadmu})--(\ref{eq:tadBmuP}).

\subsection{The scalar sector}\label{sec:ScalarsHiggsSector}
In this model, $2$ MSSM complex doublets and $2$ bilepton complex singlets are present, yielding $4$ {\it{CP}}-even, $2$ {\it{CP}}-odd, and $2$ charged physical scalars.

Concerning the {\it{CP}}-even scalars, 
the MSSM and bilepton sectors are almost decoupled, mixing exclusively due to the gauge kinetic mixing. In first approximation, the mass matrix is block-diagonal, and has mass eigenstates that mimic the MSSM case. In practice, it turns out that only two Higgs bosons are light (hereafter called $H_1$ and $H_2$, one per sector), while the other two are very heavy (above the TeV scale). The lightest scalars are well defined states, being either almost exclusively doublet-like or bilepton-like. It is worth stressing that their mixing is small (see \FIG~\ref{fig:h2mixing}) and solely due to the gauge kinetic mixing (see also \REF~\cite{Abdallah:2014fra}).

Concerning the physical pseudoscalars $A^0$ and $A^0_\eta$, their masses are given by
\begin{equation}
m^2_{A^0} = \frac{2 B_\mu}{\sin2\beta} \thickspace, \hspace{1cm} 
m^2_{A^0_\eta}  = \frac{2 B_{\mu'}}{\sin2\beta'} \thickspace.
\end{equation}
For completeness we note that the mass of charged Higgs
boson reads as in the MSSM as
\begin{equation}
m^2_{H^+} = B_\mu \left( \tan\beta+\cot\beta\right) + m^2_W\, .
\end{equation}

In this model, the {\it{CP}}-odd and charged Higgses are typically very heavy.
In eq.~(\ref{eq:tadBmu}) we see that compared to the MSSM, there is a non-negligible contribution from the gauge kinetic mixing. LHC searches limit
$\tan{\beta '} < 1.5$ and $v_x \gtrsim 7$ TeV, since~\cite{Aad:2014cka,Khachatryan:2014fba}
\begin{equation}\label{eq:Zplimit}
M_{Z'} \gtrsim 3.5~\mbox{TeV}
\end{equation}
at $95\%$ C.L.. Notice that recent reanalysis of LEP precision data also constrain $v_x \gtrsim 7$~TeV at $99\%$ C.L.~\cite{Cacciapaglia:2006pk}. A consequence of this strong constraint in the \BLSSM is that the first terms in eqs.~(\ref{eq:tadBmu})--(\ref{eq:tadBmuP}) can be large, pushing for {\it{CP}}-odd and charged Higgs masses in the TeV range.

The very large bound on the $Z'$ mass is in contrast with the non-SUSY version of the model, where the gauge couplings are free parameters and can be much smaller, hence yielding lower mass bounds. The latter need to be evaluated as a function of both gauge couplings~\cite{Basso:2012ux}.


Next, we describe the sneutrino sector, that shows two distinct features
compared to the MSSM. Firstly, it gets enlarged by the
 superpartners of the right-handed neutrinos.
 Secondly, even more drastically, a splitting between the real and imaginary
parts of each sneutrino occurs resulting in twelve states: six scalar
sneutrinos and six pseudoscalar ones
\cite{Hirsch:1997vz,Grossman:1997is}. The origin of this splitting
is the $Y^{ij}_x\,\hat{\nu}^c_i\,\hat{\eta}\,\hat{\nu}^c_j$ term in the
superpotential, eq.~(\ref{eq:superpot}), which is a $\Delta L=2$ operator
after the breaking of $U(1)_{B-L}$.  
In the case
of complex trilinear couplings or $\mu$-terms, a mixing between the
scalar and pseudoscalar particles occurs, resulting in 12 mixed states
and consequently in a $12\times 12$ mass matrix.  

\begin{figure}[hbt]
\includegraphics[width=0.9\linewidth]{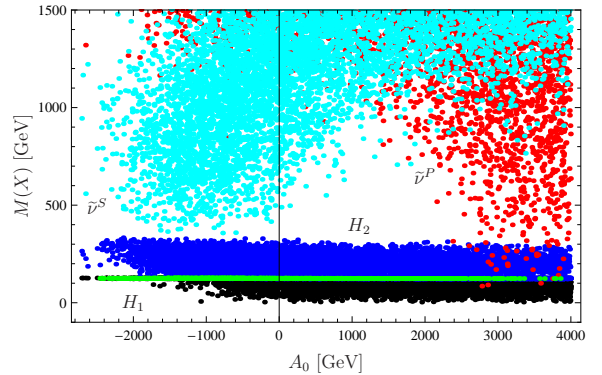}
\caption{Masses of {\it{CP}}-even ($\widetilde{\nu}^S$, cyan) and {\it{CP}}-odd ($\widetilde{\nu}^P$, red) \rsnus as a function of $A_0$. For comparison, also the masses of the lightest ($H_1$, black) and next-to-lightest ($H_2$, blue) Higgs bosons are shown. In green, it is shown configurations when $M_{H_1}=125$ GeV.
}
\label{fig:sneumasses}
\end{figure}

To gain some feeling for the behaviour of the sneutrino masses we
can consider a simplified setup: neglecting kinetic mixing as well as
left-right mixing, the masses of the \rsnus at the SUSY scale can be expressed
as
\begin{align}
\mReSnuSq \simeq & \,\, m_{\nu^c}^2
 + \mZpSq \left( \frac{1}{4} \cos(2 \beta')
                 + \frac{2 Y_x^2}{\gBL{2}} \sin\beta'^2 \right)\nonumber \\ \label{eq:mSnuA} 
 & \, \,  
 + \mZp \frac{\sqrt{2} Y_x}{\gBL{}}
    \left(A_x \sin\beta'-\mu' \cos\beta' \right)\, ,\\ 
\mImSnuSq \simeq & \,\, m_{\nu^c}^2
 + \mZpSq \left( \frac{1}{4} \cos(2 \beta')
                 + \frac{2 Y_x^2}{\gBL{2}} \sin\beta'^2 \right)\nonumber \\ 
 & \, \,  
 - \mZp \frac{\sqrt{2} Y_x}{\gBL{}}
    \left(A_x \sin\beta'-\mu' \cos\beta' \right)\, .
\label{eq:mSnuB}
\end{align}
In addition, we treat the parameters $A_x$, $m_{\nu^c}^2$, $\mZp$, $\mu'$,
$Y_x$ and $\tan\beta'$ as independent. 
The different effects on
 the sneutrino masses can easily be understood by inspecting \EQS~(\ref{eq:mSnuA})
and (\ref{eq:mSnuB}). The first two terms give always a positive
contribution whereas the third one gives a contribution that can be potentially large which differs in sign between the scalar and pseudoscalar states, therefore inducing a large mass splitting between the states. Further, this contribution can either be positive
or negative depending on the sign of 
$A_x \sin\beta'-\mu' \cos\beta'$. For example choosing 
$Y_x$ and $\mu'$ positive, one finds that the {\it{CP}}-even ({\it{CP}}-odd)
sneutrino is the lightest one for $A_x < 0$ ($A_x > 0$). This is pictorially shown in \FIG~\ref{fig:sneumasses}, as a function of the GUT-scale input parameter $A_0$, for a choice of the other parameters. One notices that the {\it{CP}}-even ({\it{CP}}-odd) sneutrino is the lightest one when the $125$ GeV Higgs boson is predominantly $H_1$ ($H_2$). It is worth pointing out here that, as will be described in the following section, when $M_{H_1}=125$ GeV, the next-to-lightest Higgs boson can decay into pairs of {\it{CP}}-even sneutrinos, but not into the similar channel with {\it{CP}}-odd sneutrinos. Being $H_2$ predominantly a bilepton field, when this decay is open it saturates its BRs, see \FIG~\ref{fig:h2BR}. Regarding the decay into {\it{CP}}-odd sneutrinos, this channel is accessible (\IE $\widetilde{\nu}^P$ is light enough) only in the region where $H_2$ is the SM-like Higgs boson, \IE mainly coming from the doublets. In this case however, this decay channel is mitigated by the small scalar mixing and is not overwhelming (unlike for $H_1$, now mainly from the bileptons). 

Depending on the parameters, either type of sneutrinos can get very light. If the LSP, it can be a suitable dark matter candidate~\cite{Basso:2012gz} and yield extra fully invisible decay channels to the Higgs bosons, thereby increasing their invisible widths. In the case of the decay into the {\it{CP}}-odd sneutrino, since this can happen mainly for the SM-like Higgs boson, one should account for the constraints on the former~\cite{Khachatryan:2014jba}.
 Eventually, the \rsnus could also get tachyonic or develop dangerous $R$-parity-violating \vevs. While the first possibility is taken into account in our numerical evaluation by \SPheno, and such points are excluded from our scans, the second case will be reviewed in the following subsection.


The last important sector for considerations that will follow is the one of the charged sleptons. See \REF~\cite{Basso:2012tr} for further details.
New SUSY breaking D-term contributions to the masses appear, that can be parametrised as a function of the $Z'$ mass and of $\tan\beta'$ as
\begin{equation}\label{B-L-Dterms}
\frac{Q^{B-L}}{2} \frac{M_{Z'} (\tan^2\beta' -1)}{1 + \tan^2 \beta'}.
\end{equation}
Their impact is larger for the sleptons than for the squarks by a factor of $3$ due to the different \BL charges ($Q^{B-L}$). It is possible to vary the stau mass  
by $\pm\mathcal{O}(100)$~GeV with respect to the MSSM case while keeping the impact on the squarks under control.
Having different sfermion masses in the \BLSSM as compared to the MSSM has a net impact onto the Higgs phenomenology, in particular in enhancing the $h\gamma\gamma$ coupling while keeping unaltered the SM-like Higgs coupling to gluons. As described at the end of this review, the new D-terms coming from the \BL sector can further reduce the stau mass entering in the $h\gamma\gamma$ effective interaction (while ensuring a pole mass of $\sim 250$ GeV, compatible with exclusions)~\footnote{With pole mass we denote the one-loop corrected mass at $Q=M_{SUSY}=\sqrt{\tilde{t}_1\tilde{t}_2}$, while in the loop, leading to the effective $h\gamma\gamma$ coupling, the running $\overline{\text{DR}}$ tree-level mass at $Q=m_h$ enters, being $h$ the SM-like Higgs boson,\IE $m_h=125$ GeV.}
 leading this mechanism to work also in the constrained version of the model. This mechanism has been recently reanalysed also in \REF~\cite{Hammad:2015eca} in the very same model.

\subsection{The issue of R-Parity conservation}

We have encountered so far several neutral scalar fields with could develop a \vev, beside the Higgs bosons. If \vevs of fields charged under QCD and electromagnetism are forbidden because the latter are good symmetries, \rsnu \vevs, which are not by themselves problematic, would unavoidably break R-Parity. The issue of conserving R-Parity is of fundamental importance, since this is a built-in symmetry in our model where \BL is gauged. We will therefore restrain ourselves to parameter configurations where the global minimum is R-Parity conserving.

When all neutral scalar fields are allowed to get a \vev, it is not trivial even at the tree level to find which is the deeper global minimum, and whether it is of a ``good'' type, here defined as having the correct broken symmetries and  being R-Parity conserving. 
One possible way to study this issue is to start from a simplified set of input parameters yielding a correct tree level global minimum when only the Higgs fields get a \vev, and then look for the true global minimum when all other neutral fields (mainly \rsnus) acquire a \vev, both at the tree level and at loop level. See \REF~\cite{CamargoMolina:2012hv} for further details.

At the tree level there seems to exist regions
where the \BLSSM has a stable, R-Parity-conserving global minimum with the correct broken and unbroken gauge groups. For this to happen one needs the \rsnu Yukawa coupling $Y_x$ to be not so large, and the trilinear parameter $A_0$ to be not large compared to the soft scalar mass $m_0$, as, intuitively,
large $Y_x$ and $A_0$ can lead to large negative contributions to the potential energy for large values of $v_x$, as well as reducing the effective \rsnu masses, as described above and clear from \FIG~\ref{fig:sneumasses}.

It turns out that when loop corrections are taken into account, few points all over such regions of parameters exist where R-Parity is not preserved anymore, or where $SU(2)_L$ or $U(1)_{B-L}$ are unbroken. This is apparently due to a very finely-tuned breaking of $SU(2)_L$ and $U(1)_{B-L}$ which often does not survive loop corrections. 
The reason for this is that besides the known large contributions of third generation (s)fermions, the additional new particles of the \BL sector also play an important role.
As previously for the charged sleptons sector, new SUSY breaking D-term contributions to the masses appear, see eq.~(\ref{B-L-Dterms}).
Since, as shown in eq.~(\ref{eq:Zplimit}), the experimental bounds require
$M_{Z'}$ to be in the multi-TeV range, 
these contributions can be much larger than in the MSSM sector, resulting in the observed importance of the corresponding loop contributions.
Furthermore, these contributions are also responsible for the restoration of $U(1)_{B-L}$ at the one-loop level.

Ultimately, overall safe regions of parameters cannot be found where the correct vacuum structure can be ensured. At the same time, if naive trends can be spotted for bad points to appear, these have nonetheless to be checked case-by-case due to the highly non-trivial scalar potential, and it might be possible that neighbour configurations still hold a valid global minimum. We will not check the validity of our scans from the vacuum point of view in the following, being confident that if any point is ruled out, a neighbour one yielding a very similar phenomenology can be found, which is allowed.

\section{A quick look to flavour observables}

Before moving to the Higgs phenomenology, we briefly show the impact on the \BLSSM model when considering the constraints arising from low energy observables. For a review of the observables as well as for the impact onto general SUSY models encompassing a seesaw mechanism, see Refs.~\cite{Abada:2014kba,Vicente:2015cka}.

\begin{figure}[hbt]
\begin{tabular}{r}
\includegraphics[width=0.76\linewidth]{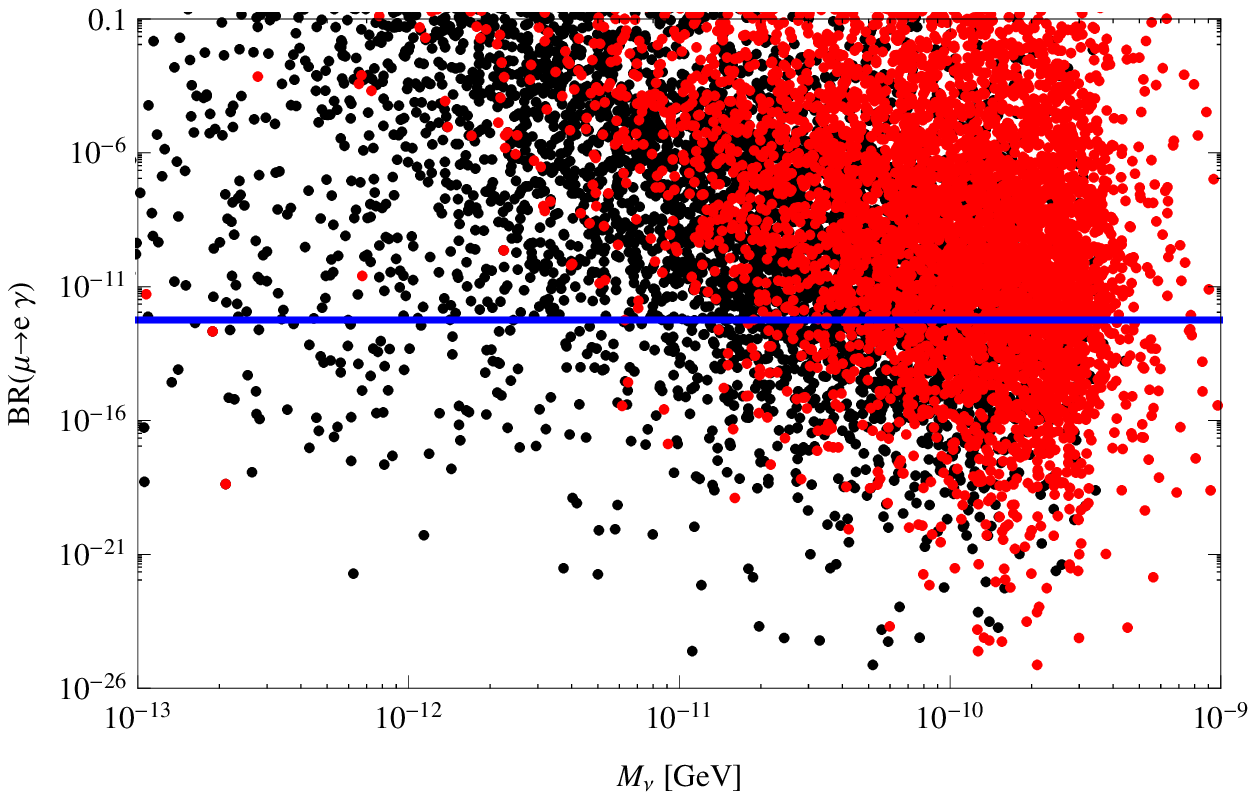} \\
\includegraphics[width=0.76\linewidth]{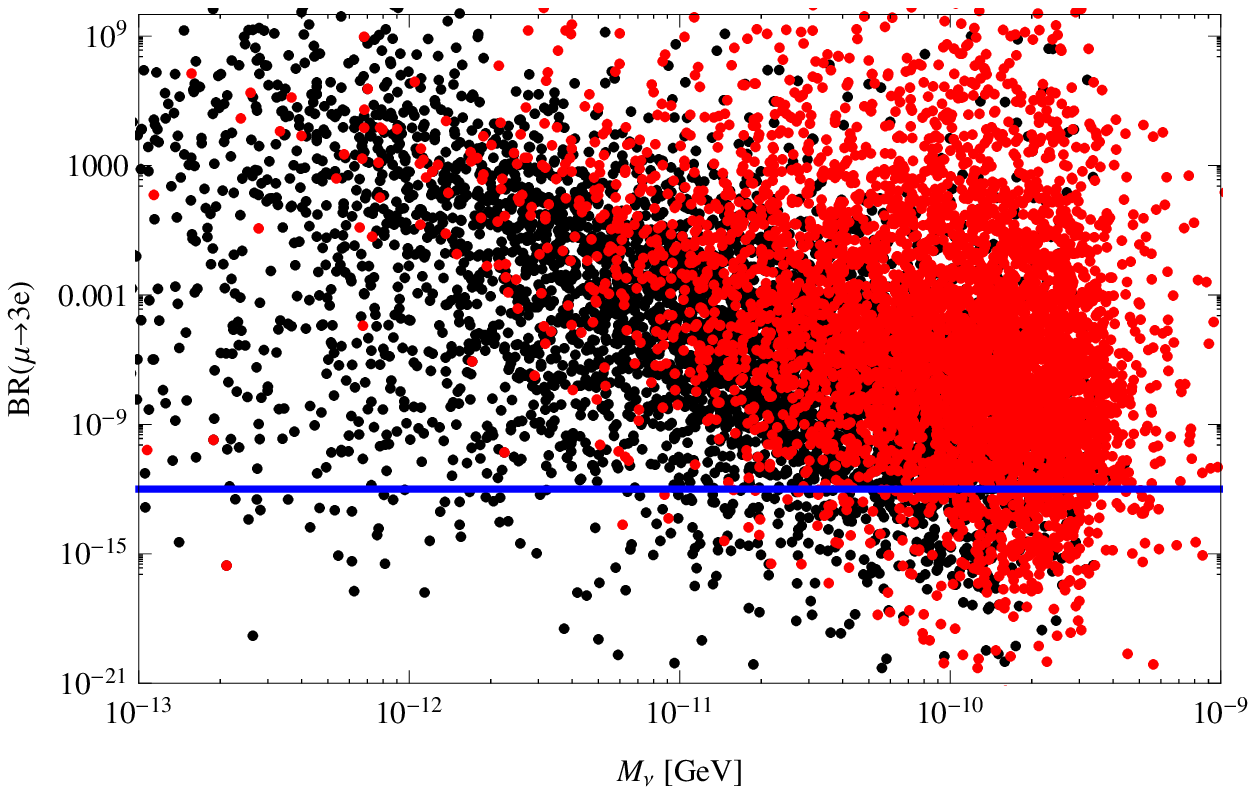} 
\end{tabular}
\caption{(Upper plot) BR($\mu \to e \gamma$) and (lower plot) BR($\mu \to 3e$) as a function of the light neutrino masses in GeV (black: $\nu_1$, red: $\nu_2$). The blue horizontal lines represent the actual experimental limits, from Refs.~\cite{Adam:2013mnn} and~\cite{Bellgardt:1987du}, respectively. The parameters have been chosen as $m_0\in [0.4,2]$~TeV, $M_{1/2} \in [1.0,2.0]$~TeV, $\tan\beta\in [5,40]$, $A_0 \in [-4.0,4.0]$ TeV, $\tan\beta ' \in [1.05,1.15]$, $M_{Z'} \in [2.5,3.5]$~TeV, $Y_x \in {\bf 1} \cdot [0.002,0.4]$, $Y_\nu \in {\bf 1} \cdot [0.05,5]\times 10^{-6}$.}
\label{fig:flavour}
\end{figure}

We consider here only the two most constraining ones, BR($\mu\to e\gamma$) and BR($\mu\to 3e$). The present exclusions are BR($\mu\to e\gamma$)~$<5.7\cdot 10^{-13}$~\cite{Adam:2013mnn} and BR($\mu\to 3e$)~$< 1\cdot 10^{-12}$~\cite{Bellgardt:1987du}. In \FIG~\ref{fig:flavour} we plot these branching ratios as a function of the mass of the lightest (in black) and next-to-lightest (in red) SM-like neutrino, which display some pattern for evading the bounds. In particular, they are required to be rather light, below $0.5$ eV, while the model, ought to the scans here performed, seems to prefer configurations with neutrinos heavier than 0.01 eV, hence the preferred region in between. Lighter mass values are nonetheless also allowed.

For convenience, the impact of satisfying the earlier bounds will be shown only in the inverted hierarchy case, due to the smaller density of configurations therein. Instead, points not allowed in the normal hierarchy case are automatically dropped.

Regarding the long-lasting $(g-2)_\mu$ discrepancy, in the setup investigated here charginos and charged Higgses are too heavy, same for the $Z'$ boson, while the neutralino and sneutrino are too weakly coupled, to give a significant enhancement over the SM prediction.

\section{Higgs phenomenology}
We review here the phenomenology of the Higgs sector, showing a first survey of its phenomenological features. First, results when normal hierarchy is imposed are presented. Then, we will show that the inverted hierarchy is also possible on a large portion of the parameter space. Without aim for completeness, the results are here presented as the starting point for a more thorough investigation.
Finally, it is described how model features pertaining to the extended gauge sector impinge onto the Higgs phenomenology, and in particular how the Higgs-to-diphoton branching ratio can be easily enhanced in this model, despite the experimental data now converging to a more SM-like behaviour than in the recent past.

\subsection{Normal hierarchy}

In this subsection we discuss the normal hierarchy case, with the lightest Higgs boson being the SM-like one (\IE, predominantly from the doublets), and a heavier Higgs boson predominantly from the bilepton fields (those carrying $B-L$ number and responsible for its spontaneous breaking). Their mixing is going to be small and solely due to the kinetic mixing.

\begin{figure}[!h]
\begin{center}
\includegraphics[width=0.5\textwidth,angle=0]{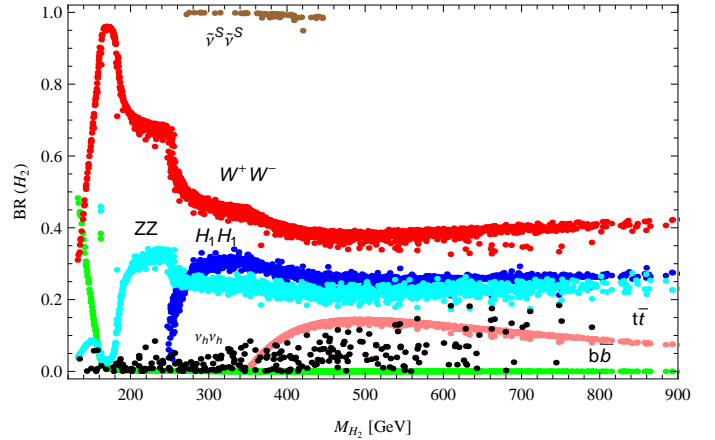}
\caption{Branching ratios for $H_2$ with $M_{H_2} > M_{H_1} = 125$ GeV. The {\it{CP}}-even sneutrino channel (brown) is superimposed. \label{fig:h2BR}}
\end{center}
\end{figure}

In \FIG~\ref{fig:h2BR} we first inspect the heavy Higgs boson branching ratios.
Besides the standard decay modes, the decay into a pair of SM Higgs bosons exist, as well as two new characteristic channels of this model, comprising right-handed (s)neutrinos.
\begin{enumerate}
\item $H_2 \to H_1 H_1$. Its BR can be up to $40\%$ before the top quark threshold, and around $30\%$ afterwards;
\item $H_2 \to \nu_h \nu_h$. A similar decay channel exists for the $Z'$ boson. The BR are $\mathcal{O}(10)\%$, up to $20\%$ depending on the heavy Higgs and neutrino masses;
\item $H_2 \to \widetilde{\nu}^S\widetilde{\nu}^S$, where, $\widetilde{\nu}^S$ is the {\it{CP}}-even sneutrino and the LSP, hence providing fully invisible decays of the heavy Higgs. If kinematically open, it saturates the Higgs BRs. Notice that only points with very light {\it{CP}}-even sneutrinos are shown, possible only for very large and negative $A_0$ (see \FIG~\ref{fig:sneumasses}).
\end{enumerate}

\begin{figure}[!h]
\begin{center}
\includegraphics[width=0.5\textwidth,angle=0]{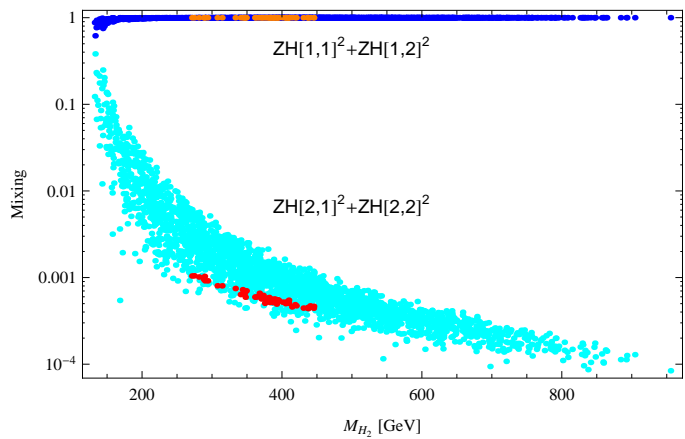}
\caption{Mixing between Higgs boson mass eigenstates (blue-orange: $H_1$, cyan-red: $H_2$) and scalar doublet fields, as a function of $M_{H_2}$. $ZH[i,j]$ is the scalar mixing matrix. Orange/red points are the subset corresponding to BR$(H_2\to \widetilde{\nu}^S\widetilde{\nu}^S) > 90\%$ .\label{fig:h2mixing}}
\end{center}
\end{figure}

\begin{figure}[!h]
\begin{center}
\includegraphics[width=0.5\textwidth,angle=0]{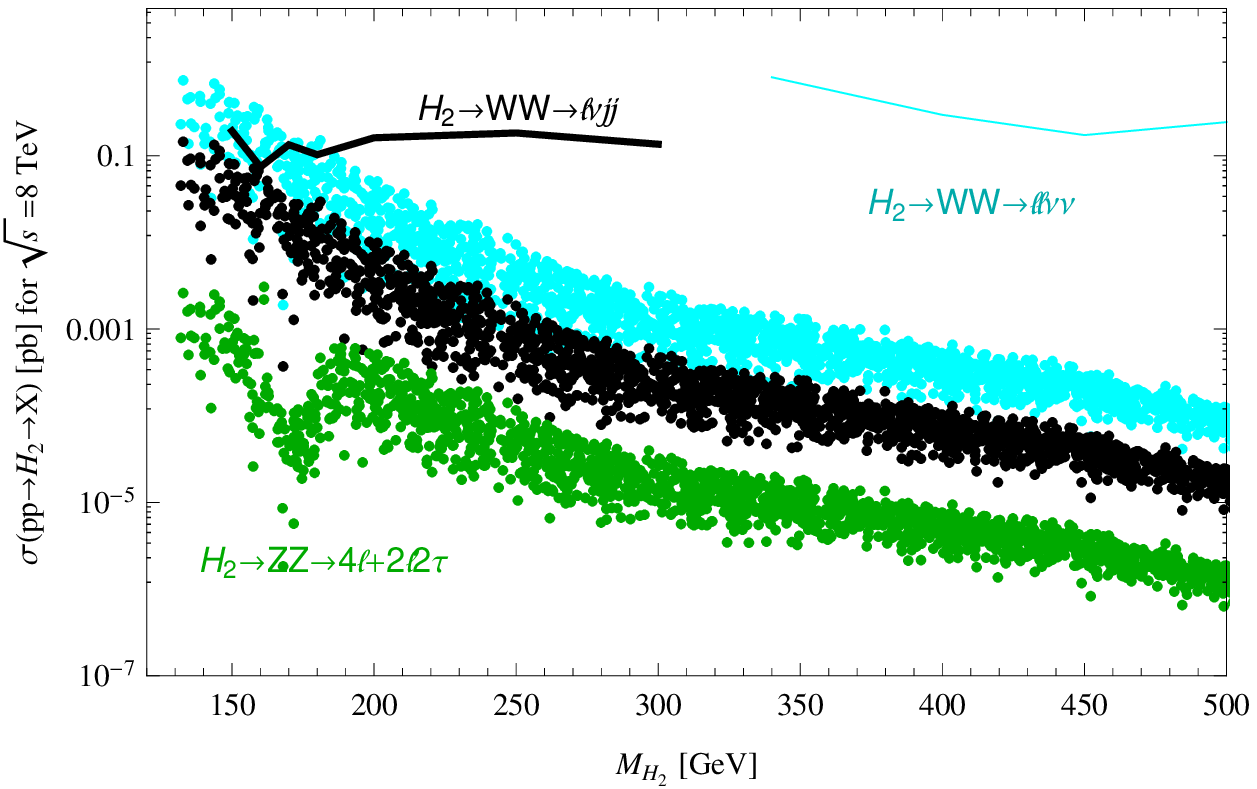}\\
\includegraphics[width=0.5\textwidth,angle=0]{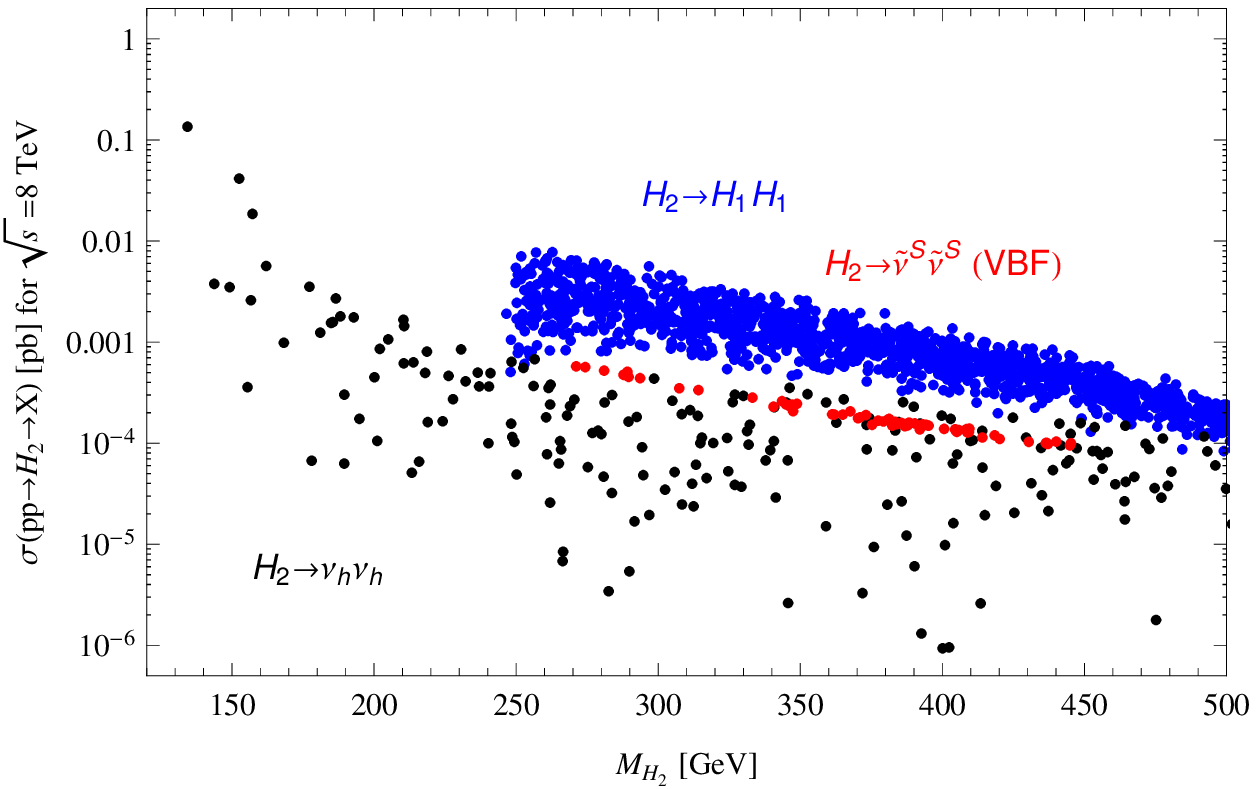}
\caption{Cross sections at $\sqrt{s} = 8$ TeV for (upper plot) the SM-like channels (lower plot) the new channels, as a function of the heavy Higgs mass. The solid lines above are the exclusion curves from~\cite{CMS-PAS-HIG-13-027,Chatrchyan:2013yoa}.
\label{fig:h2xs}}
\end{center}
\end{figure}

While the first two channels exist also in the non-SUSY version of the model~\footnote{However, in the non-SUSY \BL model the Higgs mixing angle is a free parameter, directly impacting on these branching ratios.} (see, \EG,~\cite{Basso:2010yz}), the last one, involving the {\it{CP}}-even sneutrino, is truly new and rather intriguing. This is because the sneutrino is light and it can be a viable LSP candidate if with mass lower than $H_2$, as in this case~\cite{Basso:2012gz}. It however implies that the heavy Higgs is predominantly bilepton-like, with a light Higgs very much SM-like. This can be seen in \FIG~\ref{fig:h2mixing}, where the points with large BR($H_2 \to \widetilde{\nu}^S\widetilde{\nu}^S$) (in red) have the lowest mixing between $H_2$ and the SM scalar doublet fields, of the order of $0.1\%$. It immediately follows that this channel will have very small cross section at the LHC, when considering SM-like Higgs production mechanisms.
This is true for all heavy Higgs masses $M_{H_2} > 140$ GeV. The $125$ GeV Higgs is well SM-like, with tiny reduction of its couplings to the SM particle content. On the other side, the heavy Higgs is feebly mixed with the doublets, suppressing its interactions with the SM particles, and hence its production cross section. This can be seen in \FIG~\ref{fig:h2xs} (top frame). Considering only the gluon fusion production mechanism, and multiplying it by the relevant BR, we get the cross sections for the choice of channels displayed therein. The most constraining channels, $H\to WW \to \ell\nu jj$ and $H\to WW \to 2\ell 2\nu$, are also compared to the exclusions at the LHC for $\sqrt{s}=8$ TeV from Refs.~\cite{CMS-PAS-HIG-13-027} and~\cite{Chatrchyan:2013yoa}, respectively. The $H\to ZZ$ channels are well below current exclusions, that are hence not shown. 

We see that all~\footnote{Starting from $M_{H_2} > 130$ GeV.} the displayed configurations are allowed by the current searches (the exclusions shown by solid curves of same color as the depicted channel). This is because of the suppression of the heavy Higgs boson cross sections due to the small scalar mixing.

In the lower plot are displayed the cross sections for the new channels. Those pertaining to model configurations for which the heavy Higgs boson decays to the {\it{CP}}-even sneutrino (LSP), yielding a fully invisible decay mode, are displayed in red. Contrary to the all other cases, the production of the heavy Higgs for this channel is via vector boson fusion as searched for at the LHC~\cite{CMS-PAS-HIG-14-038}. Typical cross sections range between $0.1$ fb and $1$ fb. The $H_2\to H_1H_1$ channel is shown in blue and it can yield cross sections of $1\div 10$ fb for $250 < M_{H_2} < 400$ GeV. Last is the $H_2 \to \nu_h \nu_h$ channel. It can be sizable only for very light $H_2$ masses: $\sim 10\div 100$ fb for $140 < M_{H_2} < 160$ GeV, although the further decay chain of the heavy neutrinos have to be accounted for. The latter can give spectacular multi-leptonic final states of the heavy Higgs boson ($4\ell 2\nu$ and $3\ell 2j \nu$) or high jet multiplicity ones ($2\ell 4j$), via $\nu_h \to \ell^\mp W^\pm$ and $\nu_h \to \nu Z$ in a $2:1$ ratio (modulo threshold effects). Further, these decays are typically seesaw-suppressed and can therefore give rise to displaced vertices~\cite{Basso:2008iv}.

\subsection{Inverted hierarchy}

In this subsection we discuss the inverted hierarchy case, where $H_2$ is the SM-like boson and a lighter Higgs boson exists.

\begin{figure}[!h]
\begin{center}
\includegraphics[width=0.5\textwidth,angle=0]{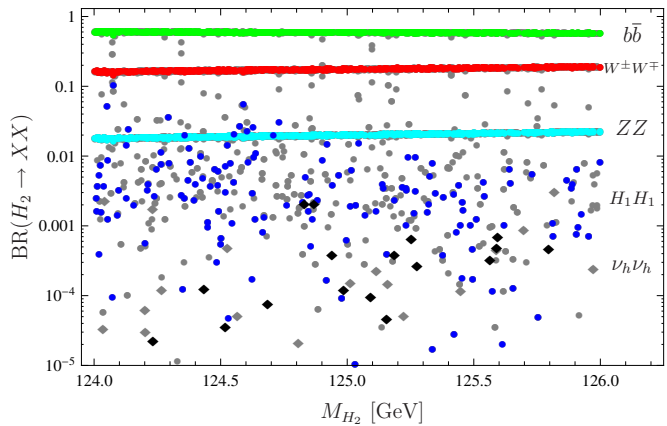}
\caption{Branching ratios for the $125$ GeV Higgs boson ($H_2$). The decay into heavy neutrinos is displayed with diamonds. All others with circles. Gray points are excluded by the low energy observables and by \HB.
The decay into {\it CP}-odd sneutrinos is not shown.
\label{fig:invH2BR}}
\end{center}
\end{figure}
We start once again by presenting the BRs for the next-to-lightest Higgs boson in \FIG~\ref{fig:invH2BR}. This time however this is the SM-like boson, hence predominantly from the doublets. It has the same new channels as the heavy Higgs in the normal hierarchy, the only difference being the {\it{CP}}-odd \rsnu instead of the {\it{CP}}-even one. This is simply because the inverted hierarchy can happen only for large positive $A_0$ values, where only the {\it{CP}}-odd \rsnu can be light, see \FIG~\ref{fig:sneumasses}. The configurations not allowed by the low energy observables or by \HB are displayed as gray points. We see that $H_2$ may have sizable decays into pairs of the lighter Higgs bosons,  yielding $4b$-jets final states. This decay is still allowed with rates up to few percent. Further, rare decays into pairs of heavy neutrinos are also present, with BRs below the permil level. This channel can give rise to rare multi-lepton/jets decays for the SM-like Higgs boson, that are searched for at the LHC, even in combination with searches for displaced vertices~\cite{Basso:inpreparation}. The last available channel is the decay into pairs of {\it{CP}}-odd \rsnus. Being the LSP, it will increase the invisible decay width and hence give larger-than-expected widths for the SM-like boson. Its rate is obviously constrained, and a precise evaluation of the allowed range is needed. It however goes beyond the scope of the present review and we postpone it to a future publication.

Regarding the lightest Higgs boson ($H_1$), this will obviously decay predominantly into pairs of $b$-jets. Notice that due to its large bilepton fraction it can also decay into pairs of very light RH neutrinos, at sizable rates depending on the neutrino masses. As in the in previous figure, the non-allowed configurations are displayed as gray points. We see that the pattern of decays is not affected by the inclusion of the constraints, in the sense that this channel stays viable. Once again, the latter will yield multi-lepton/jet final state, which will be very soft, and hence very challenging for the LHC. However, also in this case displaced vertices may appear.

\begin{figure}[!h]
\begin{center}
\includegraphics[width=0.5\textwidth,angle=0]{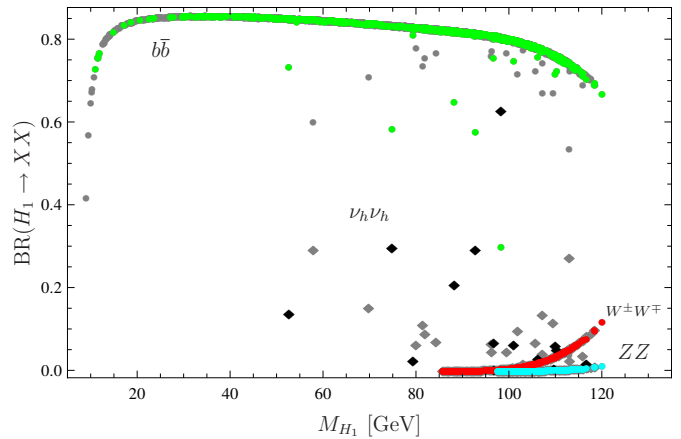}
\caption{Same as in \FIG~\ref{fig:invH2BR} for the lightest Higgs boson ($H_1$).\label{fig:invH1BR}}
\end{center}
\end{figure}

\begin{figure}[!h]
\begin{center}
\includegraphics[width=0.5\textwidth,angle=0]{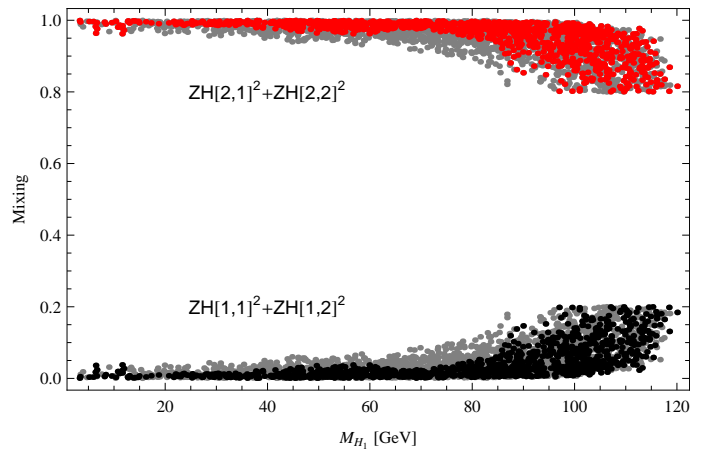}
\caption{Mixing between scalar mass eigenstates and Higgs doublets.
(black: $H_1$, red: $H_2$) and scalar doublet fields, as a function of $M_{H_1}$. $ZH[i,j]$ is the scalar mixing matrix. Gray points are excluded by the low energy observables and by \HB.
\label{fig:Hmixinv}} 
\end{center}
\end{figure}

As in the previous section, we show in \FIG~\ref{fig:Hmixinv} the mixing between the Higgs mass eigenstates and the doublet fields as a function of the light Higgs mass, to show that $H_2$ is here rather SM-like. Once more, the gray points displayed here are excluded by the low energy observables and by \HB.

\begin{figure}[!h]
\begin{center}
\includegraphics[width=0.5\textwidth,angle=0]{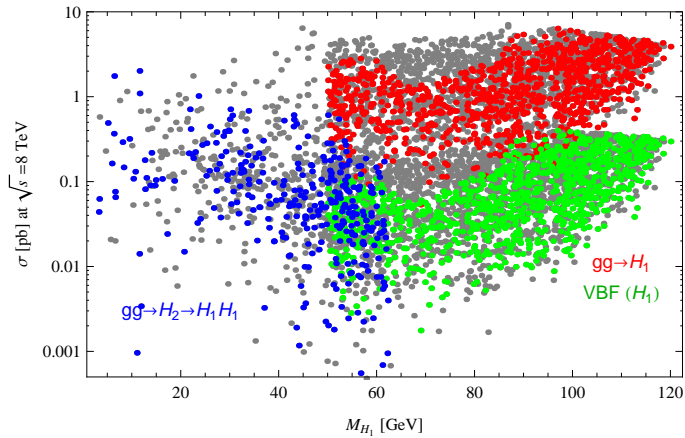}
\caption{Cross sections at $\sqrt{s} = 8$ TeV for different production mechanisms. Gluon-fusion (in red) and vector-boson-fusion (in green) mechanisms are displayed only for $M_{H_1}>50$ GeV for simplicity. Gray points are excluded by the low energy observables and by \HB.
\label{fig:h1xsinv}
}
\end{center}
\end{figure}

Finally, the production cross sections for the lightest Higgs boson can be evaluated. In \FIG~\ref{fig:h1xsinv} we compare the direct production (for the main SM production mechanisms, gluon fusion and vector boson fusion) with the pair production via $H_2$ decays only via gluon fusion, $gg \to H_2 \to H_1 H_1$. When the latter channel is kinematically open, \IE $2M_{H_1} < 125$ GeV, the lightest Higgs boson pair production has cross sections up to $1$ pb at the LHC at $\sqrt{s}=8$ TeV, and it can give rare $4b$, $2b2V$ or $4V$ ($V=W,\,Z$) decays of the SM-like Higgs boson. A thorough analysis of the phenomenology of the Higgs sector in the \BLSSM for the upcoming LHC run 2, based on the first investigations shown here, will be performed soon.

\subsection{Enhancement of the diphoton rate}

A feature of gauge-extended models is that new SUSY-breaking D-terms arise, that give further contributions to the sparticle masses. In the case of the model under consideration, we showed discussing eq.~(\ref{B-L-Dterms}) that these terms can be large, and that they bring larger corrections to sleptons than to squarks. We already discussed how the vacuum structure of the \BLSSM is affected by this. Here, we discuss their impact on the Higgs phenomenology, focusing on the Higgs-to-diphoton decay, despite disfavoured by most recent data~\cite{Khachatryan:2014jba}, as an illustrative case. See \REF~\cite{Basso:2012tr} for further details.

To start our discussion let us briefly review the partial decay width of the Higgs boson $h$ into two photons within the MSSM and its singlet extensions.
This can be written as (see, e.g.,~\cite{Djouadi:2005gi})
\begin{align}
\label{eq:decaywidth}
&\Gamma_{h \rightarrow\gamma\gamma} 
 = \frac{G_\mu\alpha^2 m_{h}^3}
{128\sqrt{2}\pi^3} \bigg| \sum_f N_c Q_f^2 g_{h ff} A_{1/2}^{h} 
(\tau_f) + g_{h WW} A_1^{h} (\tau_W) \nonumber \\ 
& \hspace*{0.2cm} + \frac{m_W^2 g_{h H^+ H^-} }{2c_W^2 
m_{H^\pm}^2} A_0^h(\tau_{H^\pm})  +  \sum_{\chi_i^\pm} \frac{2 m_W}{ m_{\chi_i^\pm}} g_{h \chi_i^+ 
\chi_i^-} A_{1/2}^{h} (\tau_{\chi_i^\pm}) \nonumber \\ 
& \hspace*{0.2cm}
+\sum_{\tilde e_i} \frac{ g_{h \tilde e_i \tilde e_i} }{ 
m_{\tilde{e}_i}^2} \,  A_0^{h} (\tau_{ {\tilde e}_i}) +\sum_{\tilde q_i} \frac{ g_{h \tilde q_i \tilde q_i} }{ 
m_{\tilde{q}_i}^2} \, 3 Q_{\tilde q_i}^2 A_0^{h} (\tau_{ {\tilde q}_i})  \bigg|^2\,,
\end{align}
corresponding to the contributions from charged SM fermions, $W$ bosons, charged Higgs, charginos, charged sleptons and squarks, respectively.
The amplitudes $A_i$ at lowest order for the 
spin--1, spin--$\frac{1}{2}$  and spin--0 particle contributions,
can be found for instance in \REF~\cite{Djouadi:2005gi}.
 $g_{hXX}$ denotes the coupling between the Higgs boson and the particle in the loop and $Q_X$ is its electric charge.
In the SM, the largest contribution is given by the $W$-loop, while the top-loop leads to a small reduction of the decay rate. In the MSSM, it is possible to get large contributions due to sleptons and squarks, although it is difficult to realise such a scenario in a constrained model with universal sfermion masses \cite{Carena:2011aa,Ellwanger:2011aa,Benbrik:2012rm}. 
In singlet or triplet extension of the MSSM also the chargino and charged Higgs can enhance the loop significantly~\cite{SchmidtHoberg:2012yy,Delgado:2012sm}. 
However, this is only possible for large singlet couplings which lead to a cut-off well below the GUT scale. In contrast, it is possible to enhance the diphoton ratio in the \BLSSM due to light staus even in the case of universal boundary conditions at the GUT scale. We show this by calculating explicitly the contributions of the stau:
\begin{align}
 & A(\tilde{\tau}) = \frac{1}{3} \frac{\partial \text{det} m_{\tilde{\tau}}^2}{\partial \log v}  \\ 
  \simeq& -\frac{2}{3} \frac{2 m_{\tau}^2 (A_\tau - \mu \tan\beta)^2}{(m_E^2 + D_R)(m_L^2 + D_L) + m_{\tau}^2 \mu \tan\beta(2 A_\tau - \mu \tan\beta)}\, .
\end{align}
Here, $D_L$ and $D_R$ represent the D-term contributions of the left- and right-handed stau and we have neglected sub-leading contributions. Given that
 $2 A_\tau < \mu \tan\beta$, for fixed values of the other parameters, $D_R$ and $D_L$ can be used to enhance the $\gamma\gamma$ rate by suppressing the denominator.

We turn now to a fully numerical analysis to demonstrate the mechanism to enhance the Higgs-to-diphoton rate as a feature of the model with an extended gauge sector. This is a result of reducing the stau mass at the Higgs mass scale via extra D-terms as shown discussing \EQ~(\ref{B-L-Dterms}). We remind here that this mechanism leaves the stop mass and hence, as we will show, the Higgs-to-gluons effective coupling nearly unchanged.
\begin{figure}[hbt]
\begin{tabular}{r}
\includegraphics[width=0.765\linewidth]{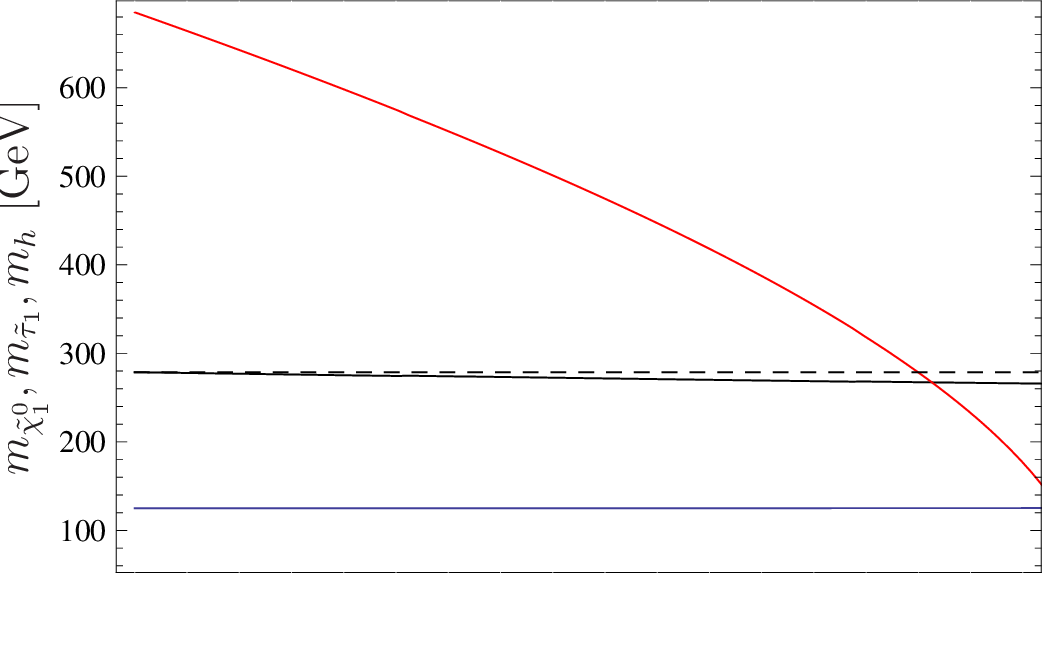} \vspace{-0.75cm}\\
\includegraphics[width=0.76\linewidth]{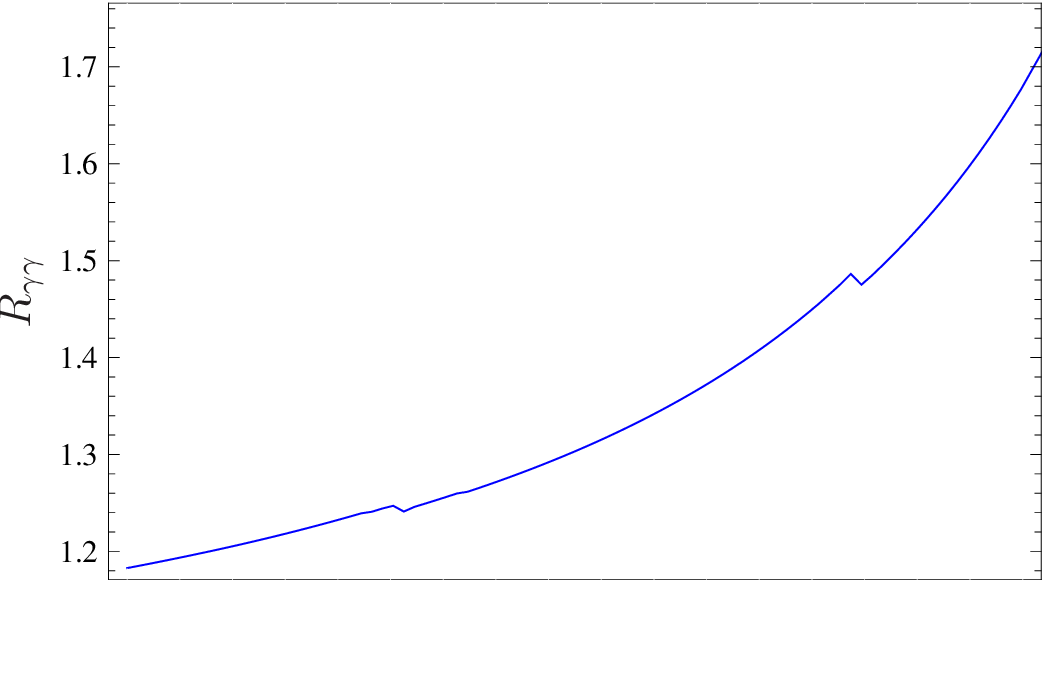} \vspace{-0.81cm}\\
\includegraphics[width=0.757\linewidth]{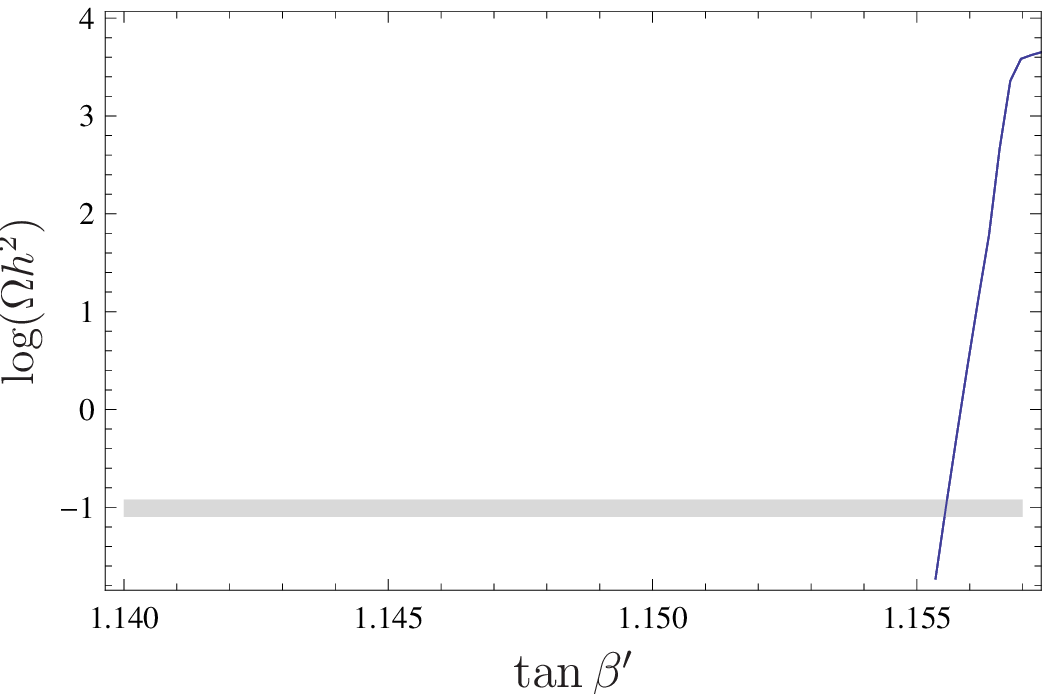} \vspace{-0.00cm}
\end{tabular}
\caption{(Top plot) The mass of the SM-like Higgs [bottom(blue line)], of the stau [middle(black) line, where the dashed line represents a reference unchanged value] and of the lightest neutralino [top(red) line]; (middle plot) the diphoton branching ratio; (bottom plot) the neutralino relic density as a function of $\tan\beta'$. The other parameters have been chosen as $m_0=673$ GeV, $M_{1/2} = 2220$~GeV, $\tan\beta=42.2$, $A_0 = -1842.6$, $M_{Z'} = 2550$~GeV, $Y_x = {\bf 1} \cdot 0.42$}
\label{fig:varTBp}
\end{figure}
In Table~\ref{tab:benchmark} we have collected two possible scenarios that provide a SM-like Higgs particle in the mass range preferred by LHC results displaying an enhanced diphoton rate. In the first point, the lightest {\it{CP}}-even scalar eigenstate is the SM-like Higgs boson while the light bilepton is roughly twice as heavy. In Fig.~\ref{fig:varTBp} we show that all the features arise from the extended gauge sector: it is sufficient to change only $\tan\beta'$ to obtain an enhanced diphoton signal $R^1_{\gamma \gamma}\equiv \frac{\left[ \sigma (gg\to h_1) \cdot BR(h_1\to \gamma\gamma)\right]_{B-L}}{\left[ \sigma (gg\to h_1) \cdot BR(h_1\to \gamma\gamma)\right]_{SM}}$ and the correct dark matter relic density while keeping the mass of the SM-like Higgs nearly unchanged. The dark matter candidate in this scenario is the lightest neutralino, that is mostly a bileptino (the superpartner of the bileptons). The correct abundance for $\tan\beta' \simeq 1.156$ is obtained due to a co-annihilation with the light stau. 
In the second point, the SM-like Higgs is accompanied by a light scalar around $98$~GeV which couples weakly to the SM gauge bosons, compatibly with the LEP excess~\cite{Barate:2003sz,Belanger:2012tt,Drees:2012fb}. In this case, the LSP is a {\it{CP}}-odd sneutrino which annihilates very efficiently due to the large $Y_x$. This usually results in a small relic density. To get an abundance which is large enough to explain the dark matter relic, the mass of the sneutrino has to be tuned below $m_W$~\cite{Basso:2012gz}. This can be achieved by slightly increasing $\tan{\beta '}$ and by tuning the Majorana Yukawa couplings $Y_x$, that tends to increase the SM-like Higgs mass for the given point. It is worth mentioning that a neutralino LSP with the correct relic density in the stau co-annihilation region can also be found in this scenario. 
Notice that both points yield rates consistent with observations in the $WW^*/ZZ^*$ channels (measured at the LHC) (being $c_{hZZ}\sim 1$), as well as an effective Higgs-to-gluon coupling close to 1.

\begin{table}[h]
\begin{tabular}{|c|cc|}
\hline \hline
& \parbox{1.5cm}{\centering Point I} 
& \parbox{1.5cm}{\centering Point II} \\
\hline
\hline
$m_{h_1}$~[GeV] & 125.2   & 98.2    \\
$m_{h_2}$~[GeV] & 186.9   & 123.0    \\
$m_{\tilde{\tau}}$ [GeV] & 267.0 & 237.3 \\
\hline
doublet fr. [\%] & 99.5 & 8.7  \\
bilepton fr. [\%]& 0.5 &  91.3 \\
\hline
$c_{h_1 g g}$ &  0.992 & 0.087   \\
$c_{h_1 Z Z}$ &  1.001 & 0.085   \\
$c_{h_2 g g}$ &  0.005 & 0.911   \\
$c_{h_2 Z Z}$ &  0.005 & 0.921   \\
\hline
$\Gamma(h_1) $~[MeV]                            & 4.13 & 0.22   \\
$R^1_{ \gamma \gamma}$ & 1.57 &  0.085   \\
$R^1_{b \overline{b}}$ & 1.03 & 0.089  \\
$R^1_{WW^*}$ & 0.98 & 0.05  \\
\hline
$\Gamma(h_2) $~[MeV]                            & 4.8 & 3.58  \\
$R^2_{ \gamma \gamma}$ & 0.005 &  1.79   \\
$R^2_{b \overline{b}}$ & 0.006 & 0.95  \\
$R^2_{WW^*}$ & 0.01 & 0.88  \\
\hline
LSP mass  ~[GeV]           &  $253.9$    &  $82.9$   \\
$\Omega h^2 $              &  $0.10$    &  $10^{-2}$   \\
\hline
\hline                    
\end{tabular}
\caption{ The input parameter used: Point I: $m_0 = 673$~GeV , $M_{1/2} = 2220$~GeV, $A_0 = -1842$~GeV, $\tan\beta=42.2$, $\tan\beta'=1.1556$, $M_{Z'} = 2550$~GeV, $Y_x = {\bf 1} \cdot 0.42$ (neutralino LSP). Point II: $m_0 = 742$~GeV , $M_{1/2} = 1572$~GeV, $A_0 = 3277$~GeV, $\tan\beta=37.8$, $\tan\beta'=1.140$, $M_{Z'} = 2365$~GeV, $Y_x=\text{diag}(0.40,0.40,0.13)$ ({\it{CP}}-odd sneutrino LSP). $c_{SVV}$ denotes the coupling squared of the Higgs fields to vector bosons normalised to the SM values. } 
\label{tab:benchmark}
\end{table}

\section{Conclusions}
In this review I described the $U(1)_{B-L}$ extension of the MSSM, focusing in particular on the scalar sector, described in details. The fundamental role that the gauge kinetic mixing plays in this sector has been underlined.

The comparison to the most constraining low energy observables showed that a preferred region for the light neutrino masses exists to evade these bounds. Then, I presented a first systematic investigation of the phenomenology of the Higgs sector of this model, showing that both the normal hierarchy and the inverted hierarchy of the two lightest Higgs bosons are naturally possible in a large portion of the parameter space. Particular attention has been devoted to analyse the new decay channels comprising both the {\it{CP}}-even and {\it{CP}}-odd \rsnus, which are a peculiarity of the \BLSSM.
Based on these first findings, a thorough analysis of the Higgs sector in the \BLSSM at the upcoming LHC run 2 will be soon prepared. The fit of the SM-like Higgs boson to the LHC data will also be performed with \HS~\cite{Bechtle:2013xfa}.

Finally, I described how in the \BLSSM model (and in general in gauge-extended MSSM models) the Higgs-to-diphoton decay can be easily enhanced.
Despite disfavoured by most recent data, this feature is a consequence of the potentially large new SUSY-breaking D-terms arising from the \BL sector. At the same time these terms affect also the vacuum structure of the model, where naive R-Parity conserving configurations at the tree level, could develop deeper R-Parity violating global minima, or partially restore the $SU(2)_L\times U(1)_{B-L}$ symmetry at one loop. It is however possible to still find R-Parity conserving global minima on the whole parameter space, which can either accommodate an enhancement of the Higgs-to-diphoton decay or fit the most recent Higgs data.

\section*{Acknowledgments} I would like to thank S.~Moretti and C.~H.~Shepherd-Themistocleous for helpful discussions in the early stages of this work. I am also really grateful to all my collaborators, and in particular to Florian Staub. I further acknowledge support from the Theorie-LHC France initiative of the CNRS/IN2P3 and from the French ANR 12 JS05 002 01 BATS@LHC.
\bibliography{BL.bib}

\end{document}